\documentclass[12pt]{article}
\usepackage{latexsym, graphicx,cite}
\usepackage{amsmath}
\usepackage{amssymb}
\usepackage{amsthm}
\allowdisplaybreaks

\usepackage[top = 1. in,bottom = 1. in, left = 1 in, right=1 in]{geometry}

 \newcommand{\arXiv}[1]{\href{http://www.arXiv.org/abs/#1}{arXiv:#1}}
\usepackage[colorlinks=true, linkcolor=blue, bookmarks=true]{hyperref}

\makeatletter
\renewcommand\section{\@startsection {section}{1}{\z@}%
                  {-3.5ex \@plus -1ex \@minus -.2ex}
                  {2.3ex \@plus.2ex}%
                  {\normalfont\large\bfseries}}
\renewcommand\subsection{\@startsection{subsection}{2}{\z@}%
                   {-3.25ex\@plus -1ex \@minus -.2ex}%
                   {1.5ex \@plus .2ex}%
                   {\normalfont\bfseries}}
\makeatother

\newcommand{\beq}{\begin{equation}}
\newcommand{\eeq}{\end{equation}}
\newcommand{\beqarr}{\begin{eqnarray}}
\newcommand{\eeqarr}{\end{eqnarray}}
\newcommand{\ber}{\begin{array}}
\newcommand{\eer}{\end{array}}
\newcommand{\del}{\partial}

\newcommand{\dsty}{\displaystyle}
\newcommand{\te}{\theta}
\newcommand{\eps}{\varepsilon}

\begin{document}

\begin{titlepage}
\begin{flushright}
\phantom{arXiv:yymm.nnnn}
\end{flushright}
\vspace{0cm}
\begin{center}
{\LARGE\bf Klein-Gordonization:\vspace{1mm}}\\ {\large\bf mapping superintegrable quantum mechanics\vspace{1.5mm}\\ to resonant spacetimes}  \\

\vskip 10mm
{\large Oleg Evnin,$^{a,b}$ Hovhannes Demirchian$^c$ and Armen Nersessian$^{d,e}$}
\vskip 7mm
{\em $^a$ Department of Physics, Faculty of Science, Chulalongkorn University,\\
Bangkok 10330, Thailand}
\vskip 3mm
{\em $^b$ Theoretische Natuurkunde, Vrije Universiteit Brussel (VUB) and\\
The International Solvay Institutes, Brussels 1050, Belgium}
\vskip 3mm
{\em $^c$ Ambartsumian Byurakan Astrophysical Observatory, Byurakan 0213, Armenia}
\vskip 3mm
{\em $^d$ Yerevan Physics Institute, 2 Alikhanyan Brothers St., 0036 Yerevan, Armenia}
\vskip 3mm
{\em $^e$ Bogoliubov Laboratory of Theoretical Physics, Joint Institute for Nuclear Research,\\ Dubna 141980, Russia}
\vskip 7mm
{\small\noindent {\tt oleg.evnin@gmail.com, demhov@yahoo.com, arnerses@ysu.am}}
\end{center}
\vspace{1cm}
\begin{center}
{\bf ABSTRACT}\vspace{3mm}
\end{center}

We describe a procedure naturally associating relativistic Klein-Gordon equations in static curved spacetimes to non-relativistic quantum motion on curved spaces in the presence of a potential. Our procedure is particularly attractive in application to (typically, superintegrable) problems whose energy spectrum is given by a quadratic function of the energy level number, since for such systems the spacetimes one obtains possess evenly spaced, resonant spectra of frequencies for scalar fields of a certain mass. This construction emerges as a generalization of the previously studied correspondence between the Higgs oscillator and Anti-de Sitter spacetime, which has been useful for both understanding weakly nonlinear dynamics in Anti-de Sitter spacetime and algebras of conserved quantities of the Higgs oscillator. Our conversion procedure (``Klein-Gordonization'') reduces to a nonlinear elliptic equation closely reminiscent of the one emerging in relation to the celebrated Yamabe problem of differential geometry. As an illustration, we explicitly demonstrate how to apply this procedure to superintegrable Rosochatius systems, resulting in a large family of spacetimes with resonant spectra for massless wave equations.

\vfill

\end{titlepage}


\section{Introduction}

Geometrization of dynamics is a recurrent theme in theoretical physics. While it has underlied such fundamental developments as the creation of General Relativity and search for unified theories of interactions, it also has a more modest (but often fruitful) aspect of reformulating conventional, well-established theories in more geometrical terms, in hope of elucidating their structure. One particular approach of the latter type is the Jacobi metric (for a contemporary treatment, see \cite{jacobi1,jacobi2,jacobi3}). This energy-dependent metric simply encodes as its geodesics the classical orbits of a nonrelativistic mechanical particle on a manifold moving in a potential.

The geometrization program we propose here can be seen as a quantum counterpart of the Jacobi metric. To a nonrelativistic quantum particle on a manifold moving in a potential, we shall associate a relativistic Klein-Gordon equation in a static spacetime of one dimension higher. Since the Klein-Gordon equation can be seen as a sort of quantization of geodesics (and reduces to the geodesic equation in the eikonal approximation), this provides a quantized version of the correspondence between particle motion on a manifold in the presence of a potential and purely geometric geodesic motion in the corresponding spacetime. Executing our geometrization algorithm in general reduces to a nonlinear elliptic equation closely reminiscent of the one emerging in relation to the Yamabe problem and its generalizations known as prescribed scalar curvature problems \cite{yamabe1,yamabe2,prescr}, and thus connects to extensive literature and interesting questions in differential geometry. (The Yamabe problem refers to constructing a conformal transformation of the given metric on a manifold that makes the Ricci scalar of the conformally transformed metric constant.)

While the correspondence we build may in principle operate on any system, we are primarily motivated by its application to a very special class of quantum systems whose energy is a quadratic function of the energy level number. Such systems are exemplified by the one-dimensional P\"oschl-Teller potential, and in higher dimensions they are typically superintegrable. In fact, our construction has been developed precisely as a generalization of the correspondence \cite{EK,EN,EN2} between the Higgs oscillator \cite{Higgs,Leemon}, a particularly simple superintegrable system with a quadratic spectrum, and Klein-Gordon equations on the Anti-de Sitter (AdS) spacetime, the maximally symmetric spacetime of constant negative curvature. This correspondence has emerged in the context of studying selection rules \cite{CEV1,CEV2,Yang} in the nonlinear perturbation theory targeting the AdS stability problem \cite{BR,review}. The correspondence has been useful for both elucidating the structure of AdS perturbation theory \cite{EN} and for resolving the old problem of constructing explicit hidden symmetry generators for the Higgs oscillator \cite{EN2}.

The reason for our emphasis on systems with quadratic spectra is that, in application to such systems, our geometrization program generates Klein-Gordon equations whose frequency spectra are linear in the frequency level number, and hence the spectrum is highly resonant (the difference of any two frequencies is integer in appropriate units). It is well-known that in the context of weakly nonlinear dynamics, highly resonant spectra have a dramatic impact, as they allow arbitrarily small nonlinear perturbations to produce arbitrarily large effects over long times. This feature has been crucial in the extensive investigations of the AdS stability problem in the literature (for a brief review and references, see \cite{review}). The main practical target of our geometrization program thus appears twofold:
\begin{itemize}
\item to provide geometric counterparts for quantum systems with quadratic spectra (the resulting Klein-Gordon equation is set up on a highly special spacetime with a resonant spectrum of frequencies and the geometric properties of this spacetime are likely to yield insights into the algebraic properties of the original quantum system, including its high degree of degeneracy and hidden symmetries),
\item to generate, starting from known quantum systems with quadratic spectra, highly resonant spacetimes (weakly nonlinear dynamics in such spacetimes is likely to be very sophisticated, sharing the features of the extensively explored weakly nonlinear dynamics of AdS).
\end{itemize}

The plan of the paper is as follows. In section 2, we formulate our general geometrization procedure and describe how it simplifies for the case of zero mass in the Klein-Gordon equation one is aiming to construct. In section 3, we describe how the previously known correspondence \cite{EK,EN,EN2} between the Higgs oscillator and AdS fits in our general framework. In section 4, we analyze the superintegrable Rosochatius system, which generalizes the Higgs oscillator, and generate a large family of spacetimes perfectly resonant with respect to the massless wave equation. We conclude with a review of the current state of our formalism and open problems.

\section{Klein-Gordon}

\subsection{General formulation of the Klein-Gordonization procedure}

Consider a quantum system with the Hamiltonian
\beq
H=-\Delta_\gamma + V(x),
\label{Hamlt}
\eeq
where $\Delta_\gamma\equiv \gamma^{-1/2}\del_i(\gamma^{1/2}\gamma^{ij}\del_j)$ is the Laplacian on a $d$-dimensional manifold parametrized with $x^i$ and endowed with the metric $\gamma_{ij}$. We shall be particularly interested in systems whose energy spectrum consists of (in general, degenerate) energy levels labelled by the level number $N=0$, 1, ..., and the energy is a quadratic function of the level number:
\beq
E_N= A(N+B)^2-C.
\label{quadrenergy}
\eeq
Such spectra are indeed observed in a number of interesting systems, typically involving superintegrability, for example:
\begin{itemize}
\item The Higgs oscillator \cite{Higgs,Leemon}, which is a particle on a $d$-sphere moving in a potential varying as the inverse cosine-squared of the polar angle.
\item The superintegrable version \cite{rossup1,rossup2} of the Rosochatius system on a $d$-sphere \cite{rosochatius,encycl}, which is the most direct generalization of the Higgs oscillator.
\item The quantum angular Calogero-Moser model \cite{FLP}.
\item The (spherical) Calogero-Higgs system \cite{HLN,CHLN}.
\end{itemize}
We additionally mention the following two completely elementary systems which give a particularly simple realization of the quadratic spectrum (\ref{quadrenergy}):
\begin{itemize}
\item A particle in one dimension in an infinite rectangular potential well.
\item The trigonometric P\"oschl-Teller system \cite{PT}. 
\end{itemize}

We would like to associate to any system of the form (\ref{Hamlt}) a Klein-Gordon equation in a certain static $(d+1)$-dimensional space-time. We introduce a scalar field $\tilde\phi(t,x)$ satisfying
\beq\label{secondorder}
-\del_t^2\tilde\phi=(-\Delta_\gamma+V(x)+C)\tilde\phi.
\eeq
In the above expression, $C$ can in principle be an arbitrary constant, but our main focus will be on systems with energy spectrum of the form (\ref{quadrenergy}) and $C$ read off from (\ref{quadrenergy}). One can equivalently write (\ref{secondorder}) as
\beq\label{KGwannabe}
\Box_{\tilde g} \tilde\phi -(C+V(x))\tilde\phi=0.
\eeq
Where $\Box_{\tilde g}$ is the D'Alembertian of the metric
\beq\label{tildeg}
\tilde g_{\mu\nu}dx^\mu dx^\nu=-dt^2+\gamma_{ij}dx^i dx^j,
\eeq
with $x^\mu=(t,x^i)$. By construction, if one implements separation of variables in (\ref{KGwannabe}) in the form
\beq
\tilde\phi=e^{i w t}\Psi(x),
\eeq
one recovers the original Schr\"odinger equation as $H\Psi=(w^2-C)\Psi$. This guarantees that the mode functions of (\ref{KGwannabe}) are directly related to the energy eigenstates of the original quantum-mechanical problem.
 Note that, if one focuses on systems with energy spectra of the form (\ref{quadrenergy}), by construction, separation of variables in (\ref{secondorder}) will lead to eigenmodes with linearly spaced frequencies:
\beq\label{omegaN}
w_N=\sqrt{A}(N+B).
\eeq
In this case, after conversion to the Klein-Gordon form, which we shall undertake below, the resulting spacetime will possess a  resonant spectrum of frequencies.

Equation (\ref{secondorder}) is not of a Klein-Gordon form, but we can try to put in this form by applying a conformal rescaling to $\tilde g$ and $\tilde\phi$:
\beq
\tilde g_{\mu\nu}=\Omega^2 g_{\mu\nu},\qquad \tilde\phi=\Omega^{\frac{1-d}2}\phi.
\eeq
One thus gets (relevant conformal transformation formulas can be retrieved, e.g., from \cite{BD})
\beq
\Box_g\phi - \left[(C+V(x))\Omega^2+\frac{d-1}2\frac{\Box_g\Omega}{\Omega}+\frac{(d-1)(d-3)}4\frac{g^{\mu\nu}\del_\mu\Omega\del_\nu\Omega}{\Omega^2}\right]\phi=0.
\eeq
If the expression in the square brackets can be made constant by a suitable choice of $\Omega$, we get a Klein-Gordon equation in a spacetime with the metric $g_{\mu\nu}$. We thus need to solve the equation
\beq\label{conformal}
(C+V(x))\Omega^2+\frac{d-1}2\frac{\Box_g\Omega}{\Omega}+\frac{(d-1)(d-3)}4\frac{g^{\mu\nu}\del_\mu\Omega\del_\nu\Omega}{\Omega^2}=m^2.
\eeq
It is wiser to rewrite this equation through the metric $\tilde g$, which is already known and given by (\ref{tildeg}):
\beq
\frac{d-1}2\Omega{\Box_{\tilde g}\Omega}-\frac{d^2-1}4\tilde g^{\mu\nu}\del_\mu\Omega\del_\nu\Omega+(C+V(x))\Omega^2=m^2.
\eeq
Since neither $V(x)$ nor $\tilde g_{\mu\nu}$ depend on $t$, one can assume that $\Omega$ is a function of $x^i$ as well. Hence,
\beq\label{confgamma}
\frac{d-1}2\Omega{\Delta_{\gamma}\Omega}-\frac{d^2-1}4\gamma^{\ij}\del_i\Omega\del_j\Omega+(C+V(x))\Omega^2=m^2.
\eeq

Note that (\ref{conformal}) is closely reminiscent of the equation emerging from the following purely geometrical problem: Consider a metric $g_{\mu\nu}$ whose Ricci scalar is $R(x)$. Is it possible to find $\Omega$ such that the Ricci scalar corresponding to $\tilde g_{\mu\nu}=\Omega^2 g_{\mu\nu}$ has a given form $\tilde R(x)$? Indeed, from the standard formulae for the change of the Ricci scalar under conformal transformations, see, e.g., (3.4) of \cite{BD}, one gets
\beq\label{rrtilde}
\Omega^2 \tilde R(x)= R(x)+2d\frac{\Box_g\Omega}{\Omega}+d(d-3)\frac{g^{\mu\nu}\del_\mu\Omega\del_\nu\Omega}{\Omega^2}.
\eeq
Algebraically, this has the same structure as (\ref{conformal}).

Equations of the form (\ref{rrtilde}) for simple specific choices of $g$ and $\tilde R$ have been studied in mathematical literature as various realizations of the `prescribed scalar curvature' problem \cite{prescr}. Substitution
\beq
\Omega=\omega^{-\frac2{d-1}},
\eeq
reduces (\ref{confgamma}) to the following compact form
\beq
-\Delta_\gamma \omega+(C+V(x))\omega=m^2\omega^\frac{d+3}{d-1},
\label{yma}
\eeq
closely reminiscent to the equation arising in relation to the Yamabe problem \cite{yamabe1,yamabe2,prescr}. (Note that the specific power of $\omega$ appearing on the righ-hand side of this equation is different from the standard Yamabe problem. This is because we are performing a conformal transformation in a spacetime of one dimension higher, rather than in the original space.) Once (\ref{yma}) has been solved, the spacetime providing geometrization of the original problem (\ref{Hamlt}) can be written explicitly as
\beq
g_{\mu\nu}dx^\mu dx^\nu=\omega^{\frac{4}{d-1}}\left(-dt^2+\gamma_{ij}dx^i dx^j\right).
\label{gsol}
\eeq

Equation (\ref{confgamma}) dramatically simplifies in one spatial dimension ($d=1$), where all the derivative terms drop out, leaving $\Omega\sqrt{C+V(x)}=m$. Thus, for the particle in an infinite rectangular potential well, Klein-Gordonization gives a massless wave equation on a slice of Minkowski space between two mirrors, while for the P\"oschl-Teller system, one immediately obtains a two-dimensional spacetime metric reminiscent of Anti-de Sitter spacetime AdS$_2$. This latter result displays some parallels to the considerations of \cite{CJP} (focusing in the hyperbolic P\"oschl-Teller system).

As we already briefly remarked, the above geometrization procedure can be applied to any Hamiltonian of the form (\ref{Hamlt}) and any $C$, irrespectively of the form of the spectrum. However, it is precisely for the spectrum and $C$ given by (\ref{quadrenergy}) that the resulting spacetime possesses the remarkable property of being highly resonant (and one may expect that its geometric properties will give a more transparent underlying pictures of the algebraic structurs of the original quantum-mechanical problem, as happens for the Higgs oscillator). We shall therefore focus on the application of our geometrization procedure to such systems with quadratic energy spectra.

\subsection{The massless case}\label{massless}

Equation (\ref{yma}) is a nonlinear elliptic equation and in general difficult to solve. Extensive existence result are established for an algebraically similar equation arising in relation to the Yamabe problem, hence one may hope that some level of understanding of solutions to (\ref{yma}) in full generality may also be attained in the future. We shall not pursue such systematic analysis here, however.

Driven by practical goals of constructing resonant spacetimes and geometrizing concrete superintegrable systems, we would like to point out  that (\ref{yma}) becomes linear and dramatically simplifies if one assumes $m^2=0$. Hence, converting a given quantum mechanical problem to a massless wave equation is considerably simpler than for general values of the mass.

We note that, if $m^2=0$, equation (\ref{yma}) looks identical to the Schr\"odinger equation corresponding to the Hamiltonian (\ref{Hamlt}), with energy eigenvalue $-C$:
\beq
-\Delta_\gamma \omega+V(x)\omega=-C\omega,
\label{yma0}
\eeq
 (Normalizable eigenstates of this energy do not generically exist, but $\omega$ does not have to satisfy the same normalizability conditions as standard wave functions, hence this should not be a problem.) Since quadratic spectra (\ref{quadrenergy}) are seen to arise from highly structured, typically superintegrable, systems, one may naturally expect that (\ref{yma0}) is amenable to analytic treatment.

There is one further assumption one might make that immediately yields solutions of (\ref{yma0}) from known solutions of the original quantum-mechanical problem (\ref{Hamlt}). Namely, imagine one has an $K$-parameter family of Hamiltonians (\ref{Hamlt}) with quadratic spectra (\ref{quadrenergy}). In this case, $A$, $B$, and $C$ are functions of the $K$ parameters defining our family of Hamiltonians. One may impose
\beq
B=0,
\eeq
which generically yields an $(K-1)$-parameter subfamily of quantum systems with quadratic spectra. Within this subfamily, the ground state $\Psi_0$ has the energy $-C$, i.e., $H\Psi_0=-C\Psi_0$. Hence, $\omega$ satisfying (\ref{yma0}) can be chosen as the vacuum state of $H$:
\beq
\omega=\Psi_0.
\label{omegaPsi}
\eeq
We shall make use of this construction below, as it allows for a straightforward application of our methodology to known exactly solvable systems. (In some cases, it is geometrically advantageous to use the non-normalizable counterpart of $\Psi_0$ with the same energy eigenvalue to define $\omega$. Such non-normalizable states should also be easy to construct for exactly solvable systems with quadratic spectra. We shall see an explicit realization of this scenario in our subsequent treatment of the superintegrable Rosochatius problem.)

As a variation of the above special case, one could force $B$ of (\ref{quadrenergy}) to be equal to a negative integer and $\omega$ to be equal to an excited state wavefunction. This, however, introduces singularities in the conformally rescaled spacetime (\ref{gsol}) at the location of zeros of the excited state wavefunctions. While one could still try to pursue this scienario by imposing appropriate constraints on the wave equation solution at the singular locus, we shall concentrate below on the most straightforward formulation (\ref{omegaPsi}) utilizing the ground state wavefunction, where the conformal factor is non-vanishing and no such subtleties arise.

\section{Higgs}

Before proceeding with novel derivations we would like to demonstrate how the case of the Higgs oscillator, which has motivated our general construction, fits into our present framework. We are essentially just reviewing the derivations in \cite{EK,EN,EN2}.

The Higgs oscillator is a particle on a sphere moving in a specific centrally symmetric potential (which we shall specify below). It is remarkable for being one of only three centrally symmetric maximally superintegrable systems on a sphere (together with free motion and the spherical Coulomb potential). A practical manifestation of superintegrability is that all of its classical trajectories are closed. The quantum version of this system has attracted considerable attention after it was reintroduced in a different guise and solved in \cite{ES}. The observed high degeneracy of energy levels of this system prompted investigation of its hidden symmetries in \cite{Higgs,Leemon}, which resulted in identification of the hidden $SU(d)$ group of symmetries for a system on a $d$-sphere, and spawned extensive literature on algebras of conserved quantities of the Higgs oscillator.  The energy spectum of the Higgs oscillator is of the form (\ref{quadrenergy}).

We shall now define, with some geometric preliminaries, the Higgs oscillator Hamiltonian. Consider a unit $d$-sphere embedded in a (d+1)-dimensional flat space as
\beq
x_0^2+x_1^2+\cdots+x_{d}^2=1
\eeq
and parametrized by the angles $\te_1$, ..., $\te_{d}$ as
\begin{align}\label{xsphere}
&x_{d}=\cos\te_{d},\qquad x_{d-1}=\sin\te_{d}\cos\te_{d-1},\\
&x_1=\sin\te_{d}\ldots\sin\te_2\cos\te_1,\qquad x_0=\sin\te_{d}\ldots\sin\te_2\sin\te_1.\nonumber
\end{align}
The sphere is endowed with the standard round metric defined recursively in $d$
\beq
ds^2_{S^d}=d\te_{d}^2+\sin^2\te_{d}ds^2_{S^{d-1}},\qquad ds^2_{S^1}=d\te_{1}^2.
\eeq
Similarly, the corresponding Laplacian is defined recursively
\beq\label{Deltasphere}
\Delta_{S^d}=\frac1{\sin^{d-1}\te_{d}}\del_{\te_{d}}\left(\sin^{d-1}\te_{d}\,\del_{\te_{d}}\right)+\frac1{\sin^2\te_{d}}\Delta_{S^{d-1}},\qquad \Delta_{S^1}=\del^2_{\te_1}.
\eeq

The Higgs oscillator is a particle on a $d$-sphere moving in a potential varying as the inverse cosine-squared of the polar angle:
\beq
H=-\Delta_{S^d}+\frac{\alpha(\alpha-1)}{\cos^2\te_d}.
\label{higgsH}
\eeq
The energy spectrum is given by
\beq\label{Higgsenrg}
E_N=\left(N+\alpha+\frac{d-1}2\right)^2-\frac{(d-1)^2}4,
\eeq
where $N$ is the energy level number. This expression is manifestly of the form (\ref{quadrenergy}).

To implement our geometrization program for the Higgs oscillator, one can work directly with (\ref{confgamma}), which takes the form
\beq
\frac{d-1}2\frac{\Omega}{\sin^{d-1}\te_d}\del_{\te_d}(\sin^{d-1}\te_d\,\del_{\te_d} \Omega)-\frac{d^2-1}4(\del_{\te_d}\Omega)^2+\left(C+\frac{\alpha(\alpha-1)}{\cos^2\te_d}\right)\Omega^2=m^2.
\eeq
Substituting $\Omega=\cos\te_d$ produces only two constraints on the parameters to ensure that the equation is satisfied:
\beq
C=\frac{(d-1)^2}4,\qquad m^2=\alpha(\alpha-1)+\frac{d^2-1}4.
\eeq
The value of $C$ above agrees with the one in (\ref{Higgsenrg}), while the relation between the Klein-Gordon mass and the Higgs potential strength is the same as found in \cite{EK}. The output of our construction is thus a family of Klein-Gordon equations on the spacetime
\beq\label{AdSHiggs}
ds^2=\frac{-dt^2+ds^2_{S^d}}{\cos^2\te_d},
\eeq
which is precisely the (global) Anti-de Sitter spacetime AdS$_{d+1}$. We note that rational values of $\alpha$ in (\ref{higgsH}) correspond to Klein-Gordon masses in AdS for which the frequency spectrum (\ref{omegaN}) is perfectly resonant (all frequencies are integer in appropriate units) rather than merely strongly resonant (differences of any two frequencies are integer in appropriate units).

A remarkable property of the Higgs oscillator is that the metric (\ref{AdSHiggs}) does not depend on the Higgs potential strength (which only affects the value of the Klein-Gordon mass). This feature is not replicated for more complicated potentials. Conversely, this implies that the AdS spacetime possesses a resonant spectrum of frequencies for fields of all masses (this statement can in fact be extended to fields of higher spins), rather than for fields of one specific mass. It is tempting to conjecture that AdS (being a maximally symmetric spacetime) is the only spacetime with this property, though we do not know a proof. Relations between Klein-Gordon equations of different masses have recently surfaced in the literature on ``mass ladder operators'' \cite{mass1,mass2,mass3,mass4}.

\section{Rosochatius}

\subsection{The superintegrable Rosochatius system}

The superintegrable Rosochatius system is the most direct generalization of the Higgs oscillator on a $d$-sphere preserving its superintegrability. General Rosochatius systems \cite{rosochatius} were among the first Liouville-integrable systems discovered. A restriction on the potential makes these systems maximally superintegrable. The Higgs oscillator can be recovered by a further restriction of the potential as a particularly simple special case. Such systems are thus an ideal testing ground for applying our machinery, which has already been shown to work for the Higgs oscillator.

The superintegrable Rosochatius systems we shall deal with here are defined by the following family of Hamiltonians:
\beq
H^{(R)}_d=-\Delta_{S^d}+\sum_{k=0}^{d}\frac{\alpha_k(\alpha_k-1)}{x_k^2}.
\label{rosH}
\eeq
The explicit form of the Laplacian and coordinates on the unit $d$-sphere can be read off from (\ref{xsphere}-\ref{Deltasphere}). The standard more general definition of the Rosochatius system \cite{rosochatius,encycl} additionally includes a harmonic potential with respect to the $x_k$ variables, $\sum_k \gamma_kx_k^2$, which gives an integrable system. If this harmonic potential is omitted, as we did above, the system becomes maximally superintegrable, as mentioned, for instance, in \cite{rossup1,rossup2}.

In order to find the spectrum of the above Hamiltonian, we shall have to apply recursively the solution of the famed one-dimensional P\"oschl-Teller problem \cite{PT}. While this material is completely standard and occasionally covered in textbooks, we find the summary given in \cite{IH} concise and convenient. The energy eigenstates of the P\"oschl-Teller Hamiltonian
\beq
H_{PT}=-\del_x^2+\frac{\mu(\mu-1)}{\cos^2x}+\frac{\nu(\nu-1)}{\sin^2x}
\label{HPT}
\eeq
are given by
\beq
\eps_n=(\mu+\nu+2n)^2,\qquad n=0,1,2,\cdots
\label{PTenergy}
\eeq
We shall not need the explicit form of the eigenfunctions satisfying $H_{PT}\Psi_n=\eps_n\Psi_n$ (though it is known).

Because of the recursion relations on $d$-spheres outlined above, the Rosochatius Hamiltonian (\ref{rosH}) can likewise be defined recursively:
\begin{align}
&H^{(R)}_d=-\frac1{\sin^{d-1}\te_{d}}\del_{\te_{d}}\left(\sin^{d-1}\te_{d}\,\del_{\te_{d}}\right)+\frac{\alpha_d(\alpha_d-1)}{\cos^2\te_d}+\frac{1}{\sin^2\te_d}H^{(R)}_{(d-1)},\\
&H^{(R)}_1=-\del^2_{\te_1}+\frac{\alpha_1(\alpha_1-1)}{\cos^2\te_1}+\frac{\alpha_0(\alpha_0-1)}{\sin^2\te_1}.
\end{align}
The variables separate, and if one substitutes the wave function in the form
\beq
\Psi(\te_1,\cdots,\te_d)=\prod_{p=1}^{d}\frac{\chi_p(\te_p)}{\sin^{(p-1)/2}\te_p},
\label{sepvar}
\eeq
one obtains a recursive family of one-dimensional eigenvalue problems, all of which are of the P\"oschl-Teller form:
\begin{align}\label{varsep}
&\left[-\del_{\te_d}^2+\frac{\alpha_d(\alpha_d-1)}{\cos^2\te_d}+\left(\frac{(d-2)^2-1}4+E_{d-1}\right)\frac1{\sin^2\te_d}-\frac{(d-1)^2}4\right]\chi_d=E_d\chi_d,\\
&\left[-\del^2_{\te_1}+\frac{\alpha_1(\alpha_1-1)}{\cos^2\te_1}+\frac{\alpha_0(\alpha_0-1)}{\sin^2\te_d}\right]\chi_1=E_1\chi_1,\nonumber
\end{align}
where $E_d$ are eigenvalues of $H^{(R)}_d$. Each subsequent equation introduces one new quantum number which we shall denote $n_d$.

The recursive solution of (\ref{varsep}) proceeds as follows. First, the solution at $d=1$ is given by (\ref{PTenergy}) as
\beq
E_1(n_1)=(\alpha_0+\alpha_1+2n_1)^2.
\eeq
At $d=2$, one gets
\beq
\left[-\del_{\te_2}^2+\frac{\alpha_2(\alpha_2-1)}{\cos^2\te_2}+\frac{(\alpha_0+\alpha_1+2n_1+\frac12)(\alpha_0+\alpha_1+2n_1-\frac12)}{\sin^2\te_2}-\frac{1}4\right]\chi_2=E_2\chi_2.
\eeq
Hence,
\beq
E_2(n_1,n_2)=\left(\alpha_0+\alpha_1+\alpha_2+2n_1+2n_2+\frac12\right)^2-\frac14.
\eeq
The general pattern can now be guessed as
\beq
E_d(n_1,\cdots,n_d)=\left(\alpha_0+\cdots+\alpha_d+2n_1+\cdots+2n_d+\frac{d-1}2\right)^2-\frac{(d-1)^2}4.
\label{rosenergy}
\eeq
It is straightforward to prove inductively that this expression persists under the recursion given by (\ref{varsep}). Note that (\ref{rosenergy}) is manifestly of the form (\ref{quadrenergy}). A classical version of the same construction, recursively expressing the superintegrable Rosochatius Hamiltonian through the action-angle variables has been given in \cite{rossup2}.

\subsection{Klein-Gordonization of the superintegrable Rosochatius system}

To demonstrate how the geometrization procedure we have proposed above operates, we shall now apply it to the superintegrable Rosochatius system. For the purposes of demonstration, we shall use the simplest formulation outlined in section \ref{massless}, which allows one to utilize known explicit solutions for ground state wavefunctions to construct the relevant massless Klein-Gordon (wave) equation.

The only technical input we shall need is the form of the ground state wavefunction of the P\"oschl-Teller Hamiltonian (\ref{HPT}) given by
\beq
\psi_0=\cos^\mu x\sin^\nu x.
\label{psi0}
\eeq
(This form satisfies the standard boundary conditions for physical wavefunctions only for $\mu\ge0$ and $\nu\ge0$. If not, $\mu$ must be replaced by $1-\mu$, and correspondingly for $\nu$. This is, however, completely irrelevant for our application of $\psi_0$ to construct geometrical conformal factors, and the above form, without any modifications, is perfectly suitable for our purposes.) From (\ref{psi0}) and the recursive construction (\ref{sepvar}-\ref{rosenergy}), one gets for the ground state wavefunction of the superintegrable Rosochatius Hamiltonian (\ref{rosH})
\beq
\Psi_0(\te_1,\cdots,\te_d)=\prod_{p=1}^d\left[\left(\cos\te_p\right)^{\alpha_p}\left(\sin\te_p\right)^{\alpha_0+\alpha_1+\cdots+\alpha_{p-1}}\right].
\eeq
On the other hand, $B$ defined by (\ref{quadrenergy}) can be read off (\ref{rosenergy}) as
\beq
B=\alpha_0+\alpha_1+\cdots+\alpha_d+\frac{d-1}2.
\eeq
We can hence directly apply the algorithm of section \ref{massless} by introducing
\beq
\omega=\prod_{p=1}^d\left[\left(\cos\te_p\right)^{\alpha_p}\left(\sin\te_p\right)^{\alpha_0+\alpha_1+\cdots+\alpha_{p-1}}\right].
\eeq
under the assumption that
\beq
\alpha_0+\alpha_1+\cdots+\alpha_d+\frac{d-1}2=0.
\eeq
This yields a $d$-parameter family of spacetimes given by (\ref{gsol}) whose massless wave equations possess perfectly resonant spectra and geometrize the superintegrable Rosochatius problem:
\beq
ds^2=\omega^{\frac{4}{d-1}}\left(-dt^2+ds^2_{S^d}\right).
\label{rosmetr}
\eeq
(Note that setting $\alpha_d=-(d-1)/2$ and the rest of $\alpha_p$ to 0 returns the case of Higgs oscillator with the coupling strength corresponding to zero mass in the Klein-Gordon equation, while (\ref{rosmetr}) becomes the AdS metric.)

For a final statement of our result, it is convinient to reparametrize $\alpha_p$ as
\beq
\alpha_p=-\frac{d-1}2\beta_p\quad\mbox{for}\quad p\ge 1,\qquad \alpha_0=-\frac{d-1}2\left(1-\beta_1-\cdots-\beta_d\right).
\eeq
In terms of $\beta_p$, (\ref{rosmetr}) becomes
\beq
ds^2=
\frac{-dt^2+ds^2_{S^d}}{\dsty\prod_{p=1}^d\left[\left(\cos\te_p\right)^{2\beta_p}\left(\sin\te_p\right)^{2(1-\beta_p-\cdots-\beta_{d})}\right]}.
\label{rosmetr_final}
\eeq
This evidently agrees with (\ref{AdSHiggs}) when $\beta_d=1$ and the rest of $\beta_p$ are zero.

\section{Outlook}

We have presented a procedure (``Klein-Gordonization'') associating to quantum systems of the form (\ref{Hamlt}) a Klein-Gordon equation on a static spacetime given by (\ref{gsol}). For systems with the quadratic energy spectrum (\ref{quadrenergy}), our procedure results in spacetimes with a resonant spectrum of evenly spaced frequencies (\ref{omegaN}). This correspondence generalizes the previously known relation between the Higgs oscillator (\ref{higgsH}) and (global) Anti-de Sitter spacetime (\ref{AdSHiggs}).

Implementing our procedure in practice requires solving a nonlinear elliptic equation, which can be written as (\ref{confgamma}) or (\ref{yma}). The latter form is closely reminiscent of elliptic equations extensively studied in relation to classic `prescribed scalar curvature' problems of differential geometry (though the exact power appearing in the power-law nonlinearity is different). If one aims at constructing a massless Klein-Gordon (i.e., wave) equation corresponding to the original quantum-mechanical system, the nonlinearity drops out, resulting in a much simpler problem. In this case, known ground state wavefunctions for the original quantum system can be utilized for the conversion procedure, as described in section \ref{massless}. We have demonstrated how this approach works for superintegrable Rosochatius systems (\ref{rosH}), resulting in a family of spacetimes (\ref{rosmetr_final}) resonant with respect to the massless wave equation.

We conclude with a list of open questions relevant for our formalism:
\begin{itemize}
\item General theory of existence of solutions of (\ref{yma}) would contribute appreciably to clarifying the operation of our formalism. Similar equations arising in differential geometry \cite{prescr} have been thoroughly analyzed, hence one should expect that the situation for our equation may as well be elucidated.
\item In practical applications of our formalism, we have focused on the case of zero Klein-Gordon mass, where (\ref{yma}) greatly simplifies. Are there any general technics for solving this equation (rather than analyzing the existence of solutions) for non-zero masses (at least, for solvable potentials in the original quantum-mechanical system).
\item Equation (\ref{yma}) may in principle admit multiple solutions, given that there is freedom in choosing boundary conditions, depending on which conformal transformation one allows. Singular conformal transformations may also be allowed (and they may push boundaries at finite distance off to infinity). This is in fact the case for the AdS construction starting from the Higgs oscillator. It would be good to quantify this freedom in choosing solutions of (\ref{yma}) and understand which prescriptions result in spacetimes interesting from a physical perspective.
\item Systems with quadratic spectra exist in extentions of the class of Hamiltonians we have considered here, given by (\ref{Hamlt}). For example, it is possible to include effects of monopole fields without distorting the spectrum \cite{monopole}. Klein-Gordonization is likely to generalize to such systems, resulting in Klein-Gordon equations with background gauge fields.
\item It would be interesting to understand how the spacetimes resulting from our construction, such as (\ref{rosmetr_final}), function in the context of dynamical theories of gravity. For instance, Anti-de Sitter spacetime solves Einstein's equations with a negative cosmological constant. More complicated spacetimes may require some matter fields to be supported as solutions. In the context of dynamical theories, the resonant linear spectra of our spacetimes will guarantee that weakly nonlinear dynamics of their perturbations is highly sophisticated. (Nonlinear instability of AdS, which is precisely a manifestation of such phenomena, is a broad currently active research area.)
\item What are the symmetry properties of spacetimes generated by ``Klein-Gordonization''? How do they connect to the symmetries of the original quantum-mechanical problem (and in particular hidden symmetries)? Again, for the case of the Higgs oscillator, this perspective has turned out to be fruitful, and it would be good to see how it works in more general cases.
\end{itemize}

\section{Acknowledgments}

The work of O.E.\ is funded under CUniverse
research promotion project by Chulalongkorn University (grant reference CUAASC).  O.E.\ furthermore thanks Marian Smoluchowski Institute of Physics in Krakow and support from Polish National Science Centre grant no.\ DEC-2012/06/A/ST2/00397, as well as Instituto de Fisica Teorica (IFT UAM-CSIC) in Madrid for its support via the Centro de Excelencia Severo Ochoa Program under Grant SEV-2016-0597 during collaboration visits while this work was in progress, and specifically Piotr Bizo\'n and Antonio Gonz\'alez-Arroyo for hospitality and discussions.
The  work of H.D. and A.N. was  partially supported by the Armenian State Committee of Science Grant No. 15T-1C367 and  was done within the ICTP programs NT04 and AF04. We furthermore thank the anonymous referee for providing an extremely detailed report with a number of stimulating suggestions.


\end{document}